\input epsf
\documentstyle[prd,aps]{revtex}
\begin{document}
\title{
Analytic approximations to the spacetime of a critical gravitational collapse}

\author{Richard H. Price}
\address{Department of Physics, University of Utah\\
Salt Lake City, UT 84112}
\author{Jorge Pullin}
\address{Center for Gravitational Physics and Geometry, Department of Physics\\
104 Davey Lab, The Pennsylvania State University, University Park, PA 16802}

\maketitle
\begin{abstract}

We present analytic expressions that approximate the behavior of the
spacetime of a collapsing spherically symmetric scalar field in the
critical regime first discovered by Choptuik.  We find that the
critical region of spacetime can usefully be divided into a
``quiescent'' region and an ``oscillatory'' region, and a moving
``transition edge'' that separates the two regions. We find that in each region
the critical solution can be well approximated by a flat spacetime
scalar field solution.  A qualitative nonlinear matching of the
solutions across the edge yields the right order of magnitude for the
oscillations of the discretely self-similar critical solution found by
Choptuik.

\end{abstract}

\vspace{-9.5cm} 
\begin{flushright}
\baselineskip=15pt
CGPG-96/1-1  \\
gr-qc/9601009\\
\end{flushright}
\vspace{8.5cm}

\section{Introduction}

The dynamics of a spherically-symmetric massless scalar field coupled
to general relativity can have two late-time outcomes. Either a black
hole forms or the scalar field energy eventually radiates away, leaving
spacetime flat.  By studying solutions very close to the critical case
(formation of a black hole of vanishing mass) Choptuik\cite{chop} has
found two very interesting phenomena, as yet not explained in any
fundamental way: universality and echoing.  Similarities to the
Choptuik results have been found in other contexts.  Abrahams and
Evans\cite{AandE} studied the collapse of axisymmetric vacuum
gravitational fields. Coleman and Evans\cite{CandE} have studied the
collapse of spherically symmetric radiative fluids, and Eardley and
Hirshmann \cite{EandH} studied the collapse of spherically symmetric
complex scalar fields.

``Universality'' is a feature of supercritical cases, those in which
holes form.  The dependence of the mass of the final hole on a
parameter of the initial data, appears to have a universal power-law
character and this character, as well as the numerical value of the
exponent appear to be the same for different families of data, though
not, as originally thought, for different collapsing fields
\cite{Maison,Evpriv}.  This feature of gravitational collapse would
seem, therefore, to have the same universal character ---as well as the
same mathematical appearance--- as phase transitions in statistical
mechanics.

In sharp contrast, ``echoing'' defies close comparison with any other
physical phenomenon. For solutions very close to criticality, the
evolving fields go through a sequence of cycles, each of which
duplicates the previous one, but on a radial scale smaller by around a
factor of 30, and on a time scale smaller by the same factor.  This
has been seen only in the work of Choptuik\cite{chop} and Abrahams and
Evans\cite{AandE} axially symmetric examples, in which the echoing
occurred with a scaling factor around 6, rather than 30 as in
Choptuik's spherically symmetric results.  Very recently
Gundlach\cite{Gu}, with an independent numerical approach, has
confirmed that the echoing is an exact feature of the solution.  For
fluid collapse\cite{CandE} and the complex scalar field\cite{EandH}
exact self-similar critical solutions are known, with no echoing (in
the complex scalar field case there is echoing in the phase).  It is
widely suspected that the echoing behavior has an origin related to
self-similarity, but a detailed understanding is still lacking.

In this paper we will consider only that feature of Choptuik's
solution which we consider most unusual, and surely that which is
least understood: the echoing.  Since no ``simple'' model problem is
yet know which has this feature, we focus instead on the equations of
Choptuik's physical problem to see the manner in which the echoing
solution arises. We are, then, in some sense dealing only with the
``how'' of the solution, rather than the ``why,'' in the hope that this
might be a useful step.

With this aim we will divide the spacetime into two regions where we
are able to find approximate solutions to the coupled Einstein-Scalar
field equations and we will discuss a nonlinear matching between
them. It wil turn out, somewhat surprisingly, that in the two relevant
regions, the behavior is mainly dominated by flat spacetime
evolutions. The oscillatory behavior, which {\em ab initio} appears as
a feature of the nonlinearities of general relativity, can actually be
accounted for by flat spacetime behavior. The nonlinear nature of the
theory is necessary to join the solution in the two regions, and this
join determines the scaling factor of the echoing.

The plan of this paper is as follows. In section II we present the
basic equations and discuss the results for the dynamics observed by
Choptuik.  In section III we present our approximate solutions and
discuss their regions of validity. In section IV we discuss the
relevance of these approximate solutions.

\section{Basic equations and Choptuik's solution}

The physical problem is the dynamics of a 
scalar field $\phi$ coupled to general relativity. The combined set of
Einstein-Scalar equations are,

\begin{eqnarray}
R_{\mu\nu} &=&8\pi \partial_\mu \phi \partial_\nu \phi\\
\Box{\phi} &=& 0.
\end{eqnarray}
We assume spherical symmetry for the scalar field and the spacetime,
so the metric can be written as
\begin{equation}\label{metric}
ds^2 = -\alpha^2(r,t) dt^2 + a^2(r,t) dr^2 +r^2 (d\theta^2 +\sin^2{\theta}\, 
d\varphi^2)\,.
\end{equation}
It is useful to introduce the mass function
\begin{equation}
m(r,t) =r {1-a^{-2}(r,t)\over 2}.
\end{equation} 
and the  dimensionless
``phase space" variables used by Choptuik:
\begin{equation}
X(r,t)= \sqrt{2 \pi} {r\over a} {\partial \phi \over \partial r},
\qquad
Y(r,t)= \sqrt{2 \pi} {r\over \alpha} {\partial \phi \over \partial t}.
\end{equation}
With these choices, the Einstein-scalar equations reduce to a set of
four partial differential equations,

\begin{eqnarray}
(a X)_{,t} &=& -{\alpha Y\over r} +(\alpha Y)_{,r}\\
(a Y)_{,t} &=& {\alpha X\over r} +(\alpha X)_{,r}\\ 
({m \over r})_{,r} + {m \over r^2} &=& {1\over r} (X^2+Y^2) \label{meq}\\ 
{(a \alpha)_{,r}\over a \alpha} &=& {2 a^2\over r} (X^2+Y^2).\label{alpheq}
\end{eqnarray}
The first two equations are equivalent to the Klein-Gordon equation in
curved spacetime, and the last two are the Einstein equations
determining the metric coefficients $a$ and $\alpha$ as  functions of
the energy density of the scalar field, which is proportional to
$X^2+Y^2$.

Note that the form of these equations is invariant  under the transformation
\begin{equation}
\alpha(r,t)\rightarrow\alpha(r,t)\sigma(t)\ \ \ dt\rightarrow dt/\sigma(t)\ .
\end{equation}
This corresponds to the fact that the $t$ coordinate is not completely
fixed by the form of (\ref{metric}). The degree of freedom inherent in
$\sigma(t)$ can be eliminated by fixing the function $\alpha(r,t)$ at
one value of $r$. In this paper we make the choice $\alpha(r=0,t)=1$ for
all $t$, so that the $t$ coordinate measures proper time of an
observer at the origin. 

For near critical solutions, there is a sequence of echoing
oscillations, each shorter than the previous one by a scaling factor
of around 30.  For supercritical solutions the sequence ends with
horizon formation; for subcritical solutions the echo sequence dies
out. For a perfectly critical solution, the solution studied by
Gundlach\cite{Gu}, the echoing would go through an infinite number of
cycles. In all cases, the duration of proper time is finite, and is
not highly sensitive to the way the sequence ends. From computational
results with many echoes, then, the time $t_*$ for the end of the
sequence can be estimated accurately. We adjust our time coordinate,
so that $t_*=0$. That is, our zero of $t$ is chosen to be the endpoint
of echoing.

Having fixed the zero and the scale of our $t$ coordinate, we next define the 
logarithmic coordinates 
\begin{equation}\label{logcoords}
\rho\equiv\ln{r}\ \ \ \ \tau\equiv\ln({-t})\ .
\end{equation}
Notice that our choice of $\alpha$, and hence of $t$, differs from that of
Choptuik\cite{chop} [he chooses \hbox{$\alpha(r=\infty,t)=1$}] but
$\rho,\tau$ defined in (\ref{logcoords}) agree with Choptuik's
logarithmic coordinates $\rho,\tau$.

Figure 1 shows a series of snapshots of the scalar field variable $X$
as a function of $\rho$ taken at various late times $\tau$.  These are
data made available by Choptuik on the internet from his numerical
evolutions.  In each of the plots we can distinguish two regions. On
the left of the plots there is a ``quiescent'' region in which the
solution dies off; on the right is an ``oscillatory'' region in which
the solution has a sinusoidal appearance. We call the moving boundary
between the two regions the ``transition edge.''
\begin{figure}
\epsfxsize=400pt \epsfbox{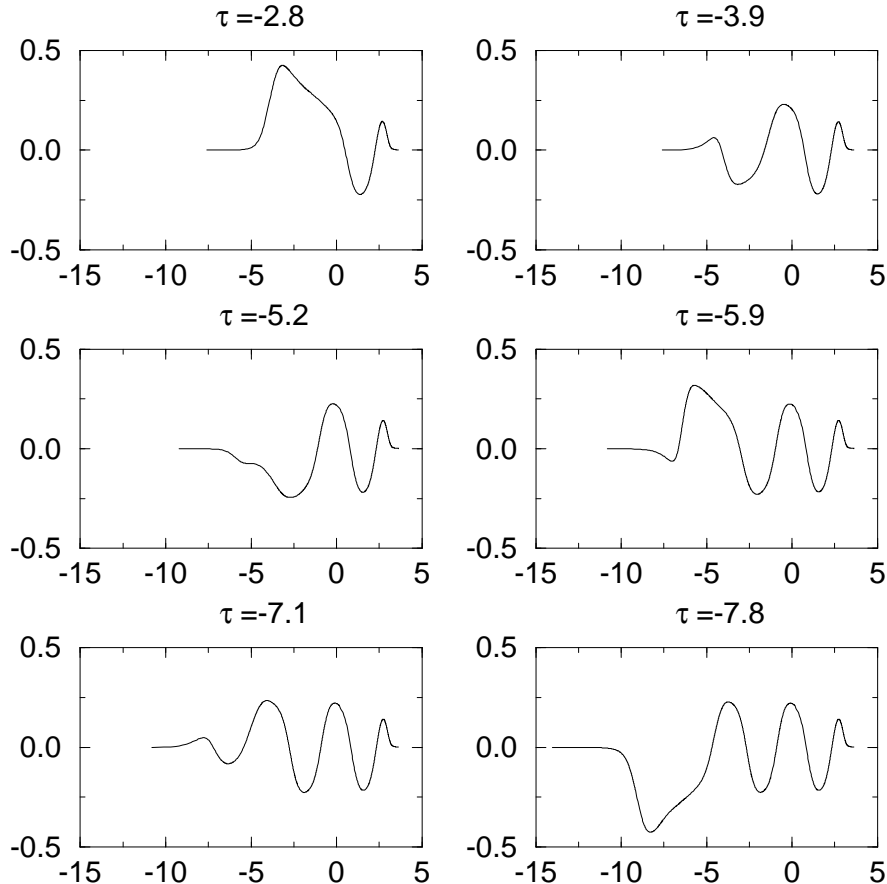}
\caption{
The Choptuik numerical results for the scalar field variable $X$ as a
function of the radial coordinate $\rho$, at different times $\tau$. }
\end{figure}

``Echoing'' of the solution is the property that the pattern of $X(r,t)$
repeats itself on a changed radial and time scale according to
\begin{equation}\label{echo1}
X(r,t)=X(e^\Delta r,e^\Delta t)\,,
\end{equation}
or equivalently
\begin{equation}\label{echo2}
X(\rho,\tau)=X(\rho+\Delta, \tau+\Delta)\,.
\end{equation}
If $\Delta$ were a continuous variable, this property would tell us
that $X$ is ``self similar,'' i.e., a function only of $r/t$.  But
equations (\ref{echo1}) and (\ref{echo2}) are valid only for a fixed
numerical value for $\Delta$, approximately 3.4 (a value of
$3.4439\pm0.0004$ according to Ref.\cite{Gu}).  For this reason the
echoing property has been referred to as a ``discrete self
similarity.''

Figure 2 shows the behavior of other variables of the problem, as
functions of $\rho$ at a fixed $\tau$.

\begin{figure}
\epsfxsize=400pt \epsfbox{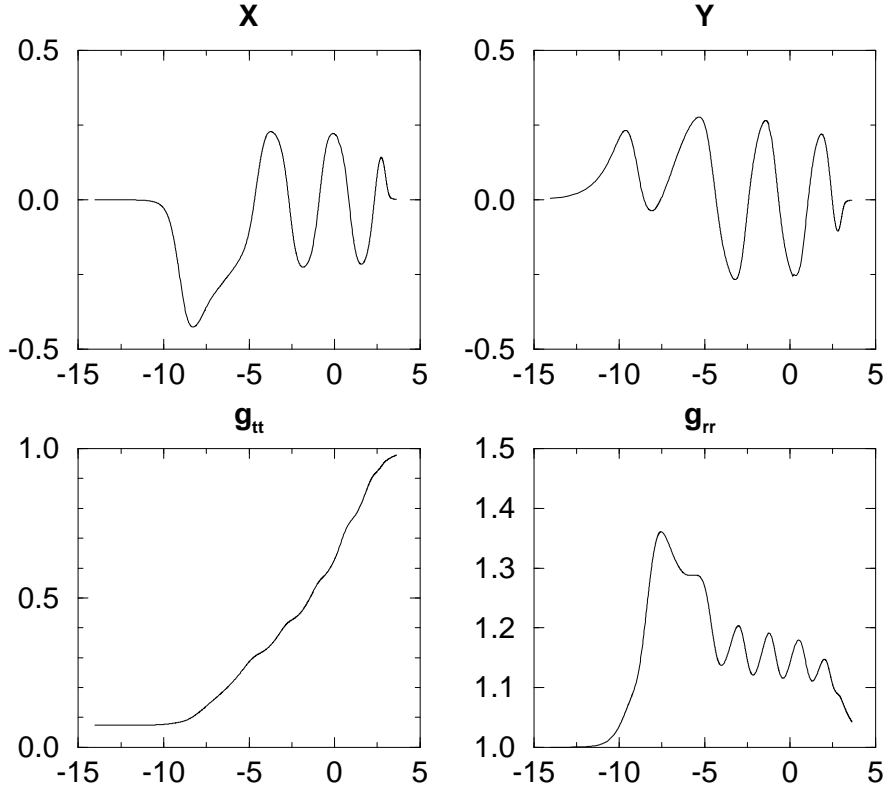}
\caption{
The Choptuik numerical results 
for scalar field and metric variables, as a function
of $\rho$, all at time $\tau=-7.87488$.}
\end{figure}
The scalar field variables $X$
and $Y$ have the appearance of being sinusoids approximately 150
degrees out of phase.  Choptuik\cite{chop2} has shown that it is
useful to view the relation of $X$ and $Y$ in phase space plots, as in
Fig.\ 3.  

The elliptical portion of the plot corresponds to the oscillatory
region of the solution. The transition region appears as the vertical
spike which connects the elliptical windings to the origin. As radius
increases, the windings decrease in size and the curve connects to the
origin at a point representing spatial infinity in the numerical
solutions. In a truly critical solution there would be an infinite
number of windings and the connection with spatial infinity would be
absent.

\begin{figure}
\epsfxsize=400pt \epsfbox{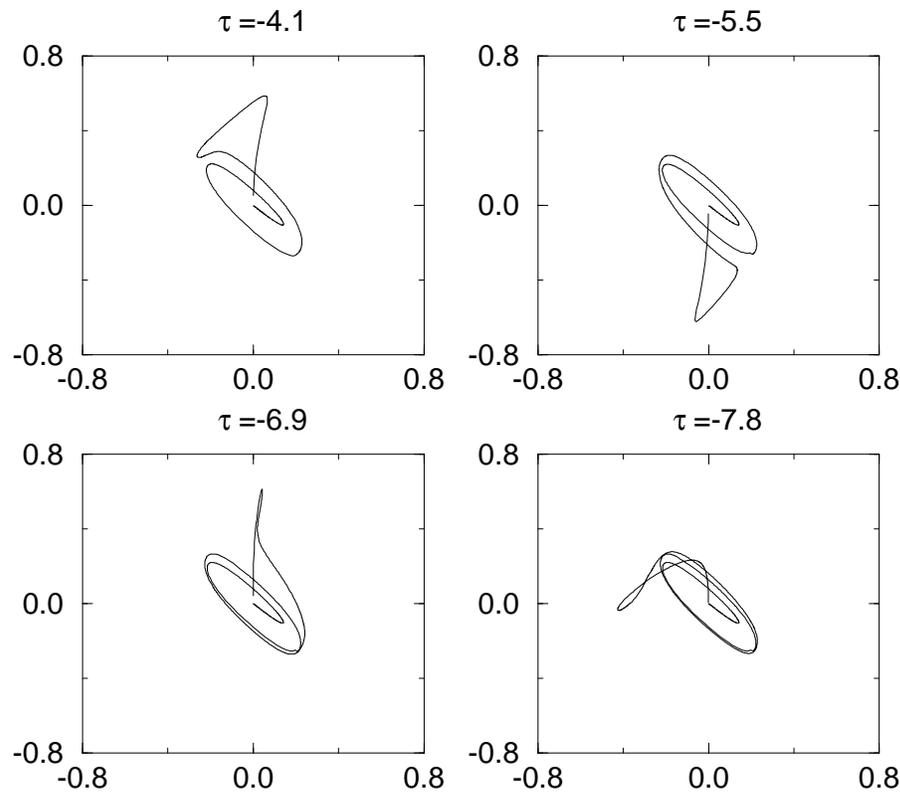}
\caption{
The Choptuik numerical results for the relationship of the phase space
variables $X$ and $Y$. Each plot corresponds to a single moment of
$\tau$ and the full range of radii.  The elliptical winding spiralling
into the origin corresponds to radius growing large; the vertical
spike approaching the origin corresponds to radius going to zero.}
\end{figure}

The qualitative features of $\alpha$, in Fig.\ 2, can be
understood from (\ref{alpheq}).  In this equation $\alpha_{,\rho}$ is
``fed'' by the source term $a^2 (X^2 +Y^2)$ which at each instant of
time is a quantity bounded between zero and a value less than one.  To
get a rough understanding of the behavior of $\alpha$ we can take the
source term as approximately constant and then the solution for
$\alpha$ is
\begin{equation}\label{alphbehavior}
\alpha\sim {\cal F}(\tau) e^{\kappa\rho}\ , 
\end{equation} 
where $k$ is a constant related to the average value of $a^2(X^2+Y^2)$ 
and  ${\cal F}$ is a function of time.

In view of the echoing, it is worthwhile to rewrite the equations in
term of a new variable $\mu=\tau - \rho$.  
\begin{figure}
${}^{}$\hspace{3cm}\epsfxsize=250pt \epsfbox{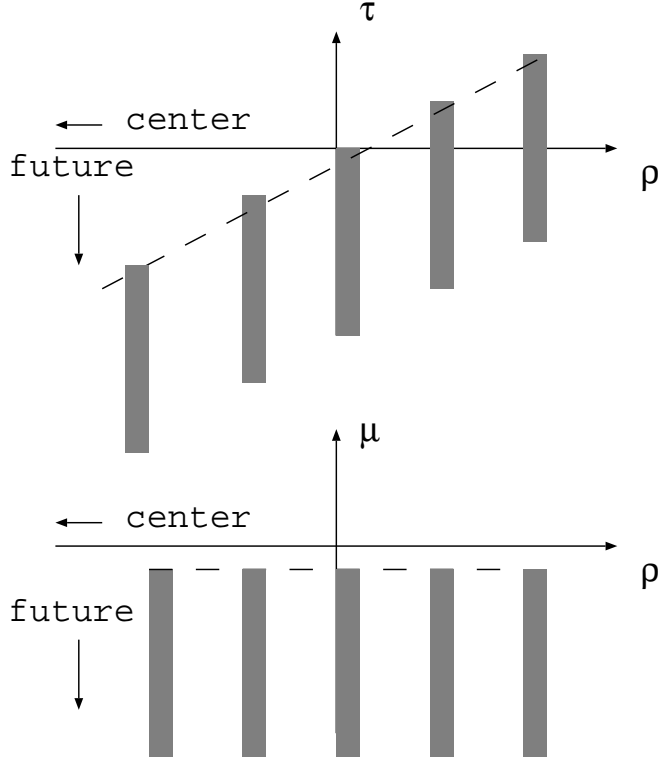}
\caption{Relationship of $\tau,\rho$ and $\mu,\rho$ coordinates. 
The dark bars represent the peaks of the $X$ oscillations; the dashed
line shows the location of the transition edge. In both coordinate
systems the region well above the transition edge is the ``quiescent''
region, and the region well below the edge is the ``oscillatory''
region.}
\end{figure}
Figure 4 shows schematically
the location of the quiescent and oscillatory regions and of the
transition edge, in terms of the $\tau,\rho$ and $\tau,\mu$
coordinates. In terms of these last coordinates, the Einstein-scalar
equations are,

\begin{eqnarray}
-e^{-\mu} (a X)_{,\mu} &=& -\alpha Y +(\alpha Y)_{,\rho}-(\alpha Y)_{,\mu},
\label{eq1}\\
-e^{-\mu} (a Y)_{,\mu} &=& \alpha X +(\alpha X)_{,\rho}-(\alpha X)_{,\mu},
\label{eq2}\\
({m\over r})_{,\rho}-({m\over r})_{,\mu}+{m\over r} &=& X^2 +Y^2
\label{eq3}\\
(\ln (a \alpha))_{,\rho}-(\ln (a \alpha))_{,\mu} &=& 2 a^2 (X^2+Y^2)\ .
\label{eq4}
\end{eqnarray}
As an example of the usefulness of the $\mu$ coordinate, we now go
back to the behavior of $\alpha$ in (\ref{alphbehavior}) and consider
the nature of ${\cal F}(\tau)$.  At any $\tau$ the change in the
``source'' $a^2(X^2+Y^2)$ becomes sizeable at the transition edge
$\mu\approx0$. This tells us that at different times the
characteristic rise of $\alpha$ as a function of $\rho$ will always
occur around $\mu\approx0$.  For this to be true, and for $\alpha$ to
have the form $e^{k\rho}$ at any one $\tau$, requires that the
dependence of $\alpha$ on $\tau$ and $\rho$ have the form
\begin{equation}
\alpha \sim e^{-\kappa\mu}
\end{equation}
in the oscillatory region, i.e., for $\mu\ll-1$.  This approximation
gives $\alpha$ a self-similar form, and is too crude, since the exact
location at which $X,Y$ start to be large is not really $\mu=0$ but
(since the solutions only have a {\em discrete} self similarity)
oscillates in $\tau$. A better approximation which takes this
oscillation into account, is
\begin{equation}
\alpha=A(\tau) e^{-\kappa\mu}\ ,
\end{equation}
where $A(\tau)$ is a bounded, oscillatory function.

\section{Approximate solutions}

\subsection{Oscillatory region $\mu<<-1$}

We make the {\em ansatz} (motivated by the results in Figs.\ 1 and 2\,)
that the $\mu$-derivative terms on the right in equations
(\ref{eq1}) and (\ref{eq2}) are to be ignored in comparison with the
$e^{-\mu}\partial_\mu$ derivatives on the left. We 
also introduce the variable
\begin{equation}
u\equiv t/r=-e^\mu
\end{equation}
in terms of which the equations take the form
\begin{eqnarray}
(a X)_{,u}  &=&-\alpha Y + (\alpha Y)_{,\rho}\label{osc1}\\
(a Y)_{,u}  &=&\alpha X + (\alpha X)_{,\rho}\label{osc2}  \ .
\end{eqnarray}
If we treat $a$ and $\alpha$
as constant, the equations  admit the solution
\begin{eqnarray}
Y &=& A \sin [\omega(\rho -\alpha u)]\label{simple1}\\
X &=& A \sin [\omega(\rho -\alpha u) -\delta]\label{simple2}
\end{eqnarray}
with,
\begin{eqnarray}
1+{1\over \omega^2} &=&a^2\\
\cos \delta &=& -{1\over a}\\
\sin \delta &=& {1\over \omega a}\ .
\end{eqnarray}
If one fits this solution to the Choptuik's data in Figs.~1 and 2, one
finds that $\delta\approx150$ degrees. The ratio of side lengths of
the phase space ellipse is given by $\tan{(\delta/2)}$, which yields a
value of approximately $4$, agreeing with the data due to Choptuik
exhibited in Fig.\ 3.

Since we have assumed that $a$ and $\alpha$ are constant, we have in effect 
been considering the spacetime to be a flat background. This suggests that we
should be able to find similar behavior in a solution to the scalar field
wave equation in Minkowski spacetime. In this case $\phi$
can always be expressed, in terms of arbitrary functions $f,g$, as
\begin{equation}\label{flatphi}
\phi={f(t-r)+g(t+r)\over r}\ .\\
\end{equation}
In terms of $f$ and $g$, Choptuik's phase space variables $X,Y$ are
\begin{eqnarray}
Y&=&f'(t-r)+g'(t+r)\label{flatY}\\
X&=&-{1\over r} (f(t-r)+g(t+r)) -f'(t-r)+g'(t+r)\label{flatX}\ .
\end{eqnarray}
To match a solution of this form to that of (\ref{simple1}),
(\ref{simple2}) we take $f$ and $g$ to have forms chosen to give good
agreement with (\ref{simple1}), (\ref{simple2}) for $\mu\ll-1$;
\begin{eqnarray}
f(\eta)&=&A\frac{1+k}{2}\frac{\eta}{1+\omega^2}\left[
\sin{(\omega\ln{\beta\eta})}-\omega\cos{(\omega\ln{\beta\eta})}
\right]\label{flata}\\
g(\eta)&=&A\frac{1-k}{2}\frac{\eta}{1+\omega^2}\left[
\sin{(\omega\ln{\beta\eta})}-\omega\cos{(\omega\ln{\beta\eta})}
\right]\ .\label{flatb}
\end{eqnarray}
Here $A$ and $\beta$ are adjustable constants
and $k^2=1+1/\omega^2$. The agreement of this
solution with the numerical results of Choptuik is most impressive in
a phase-space plot, as in Fig.\ 5, where we have used the values
$A=0.24, \omega=1.85, k=1.137$ in the flat spacetime solution.
(The value of $\beta$ does not influence the appearance of the
solution in this plot.)

\begin{figure}
${}^{}$\hspace{2cm}\epsfxsize=250pt \epsfbox{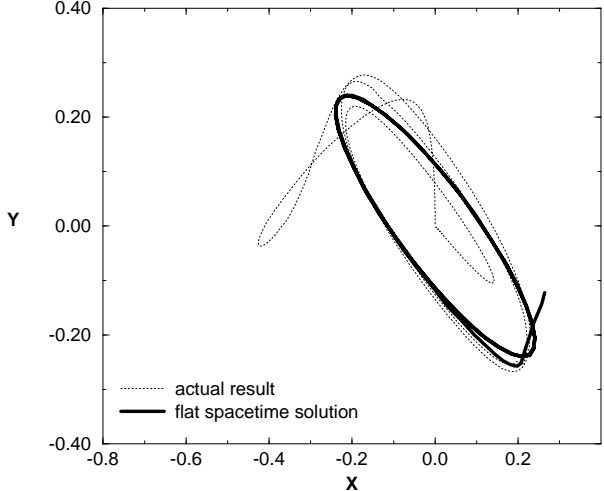}
\caption{
A phase-space plot comparing Choptuik's numerical results with a flat
spacetime solution at $\tau=-7.87488$. The flat spacetime solution
corresponds to parameters $\omega=1.85$ and $A=0.24$.}
\end{figure}
The light dotted curve is Choptuik's result for $\tau=-7.87488$ and the
thick curve is the flat spacetime solution. The thick curve,
of course, lacks the vertical spike representing the transition
edge. For the flat spacetime solution there is no transition
front. The solution in (\ref{flata}),(\ref{flatb}) is a valid solution
except at $t+r=0$ or $t-r=0$ (and is shown in Fig.\ 5 only for $t-r<0$
and $t+r>0$). It can agree with the Choptuik numerical results only in
the region in which the approximations (\ref{osc1}),(\ref{osc2})
apply, i.e., it can agree only in the oscillatory region.

The flat spacetime fit of eqs.~(\ref{flata}),(\ref{flatb}), is plotted
and compared to the numerical results of Choptuik in Figs.\ 6a,b.
 The agreement is
remarkable in the oscillatory region,  except at large $\rho$ (where
Choptuik's results deviate from the ideal critical solution). The
flat spacetime solution stops being reasonably accurate, of course,
near the transition front, and deviates wildly from Choptuik's results
in the quiescent region.

Despite the impressive agreement (in the appropriate region) of the
simple solution above with the numerical results, the solution is less
than ideal. For one thing it does not explain the gradual decrease in
the amplitude of the oscillations seen in the numerical results. More
worrisome is the question whether it is justified to treat $\alpha$ as
constant when Fig.\ 2 shows it to be increasing exponentially. An
improved approximation scheme which deals with such issues is
possible, but is not central to our main point of this paper.

\begin{figure}
${}^{}$\hspace{2cm}\epsfxsize=250pt \epsfbox{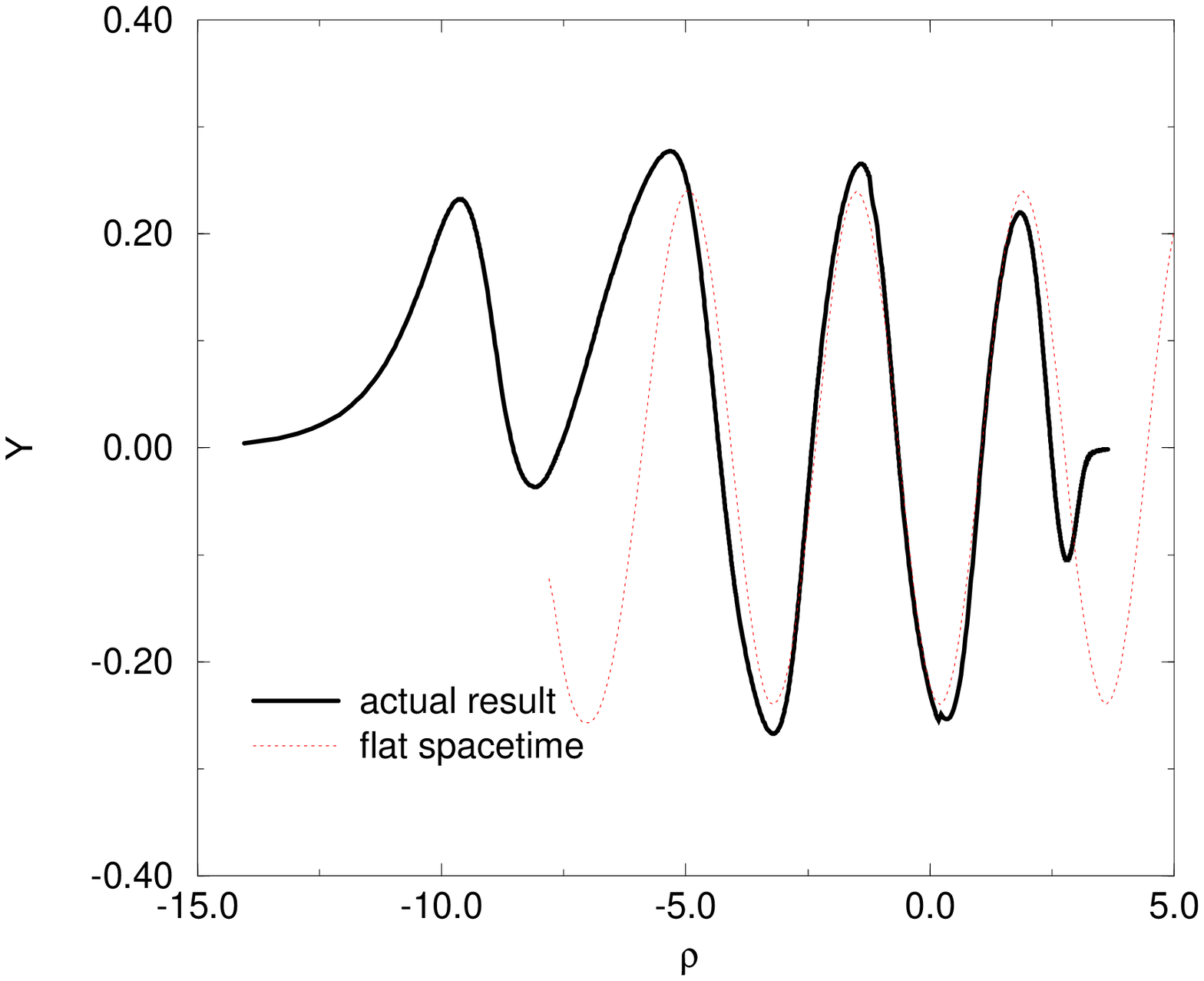}
 
${}^{}$\hspace{2cm}\epsfxsize=250pt \epsfbox{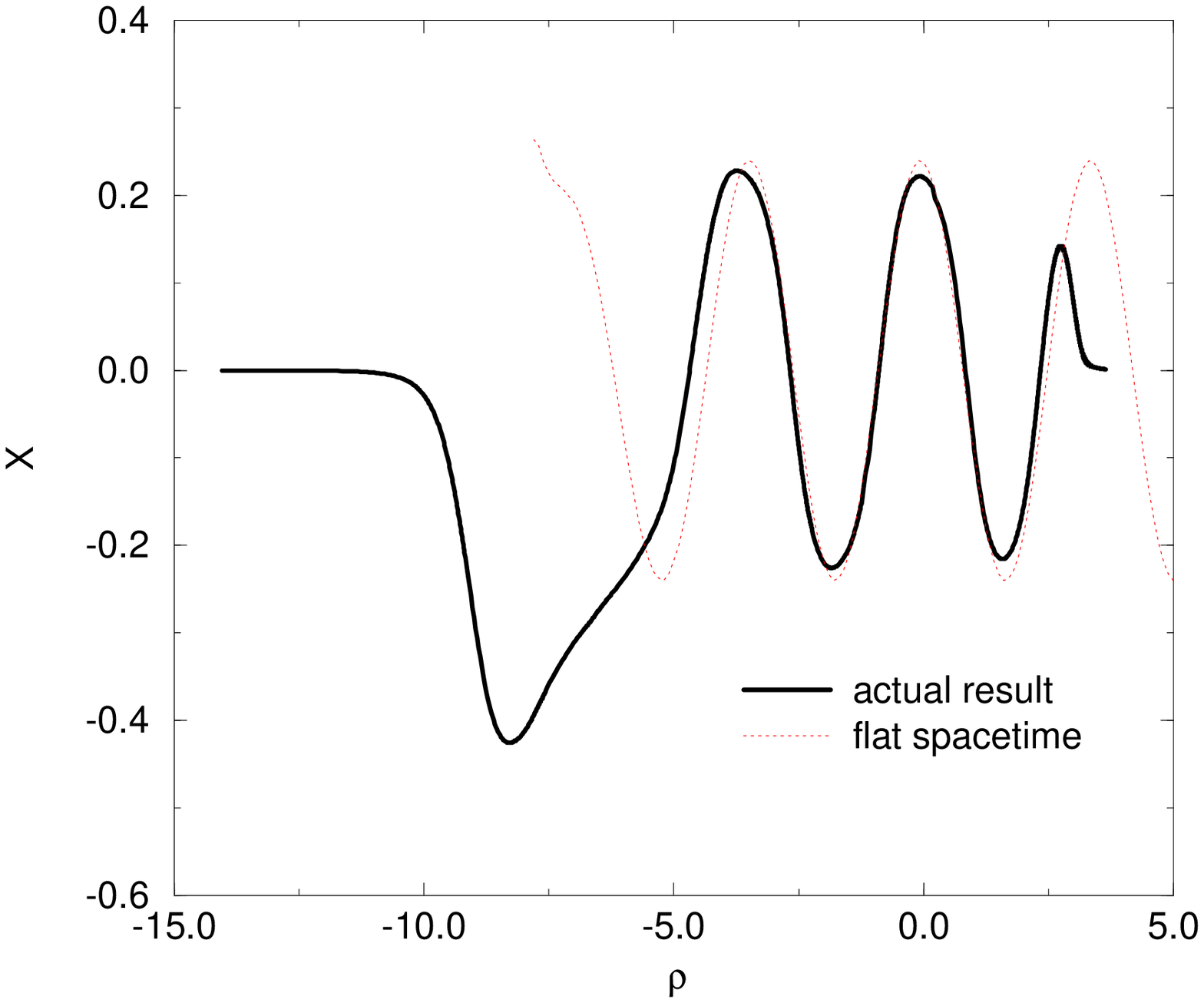}
\caption{
The solutions $X(\rho)$, $Y(\rho)$ found by Choptuik for
$\tau=-7.87488$ are compared to the flat spacetime solutions
corresponding to parameters $\omega=1.85, A=0.24, \beta=0.35$}
\end{figure}

\subsection{Quiescent region $\mu>>-1$}

In this region we drop the terms involving $e^{-\mu} \partial_{\mu}$
on the left of (\ref{eq1}) and (\ref{eq2}), and keep the
$\partial_{\mu}$ terms.  We make the convenient definitions
$x\equiv\alpha X$ and $y\equiv\alpha Y$ so that the equations for $X$
and $Y$ (with the $e^{-\mu} \partial_{\mu}$ terms dropped) read,
\begin{eqnarray}
y_{,\rho} -y_{,\mu}&=&y\\
x_{,\rho} -x_{,\mu}&=&-x\ .
\end{eqnarray}
The general solution to these equations  can immediately be written as
\begin{equation}
y = e^{(-\mu+\rho)/2} F(\tau)\ \ \ \
x =e^{(\mu-\rho)/2} G(\tau)\ ,
\end{equation}
in which $F$ and $G$ are arbitrary functions.
At $\mu=0$ these solutions become
\begin{equation}
y =  e^{\rho/2} F(\rho)\ \ \ \
x = e^{-\rho/2} G(\rho)\ .
\end{equation}
\begin{figure}
${}^{}$\hspace{2cm}\epsfxsize=250pt \epsfbox{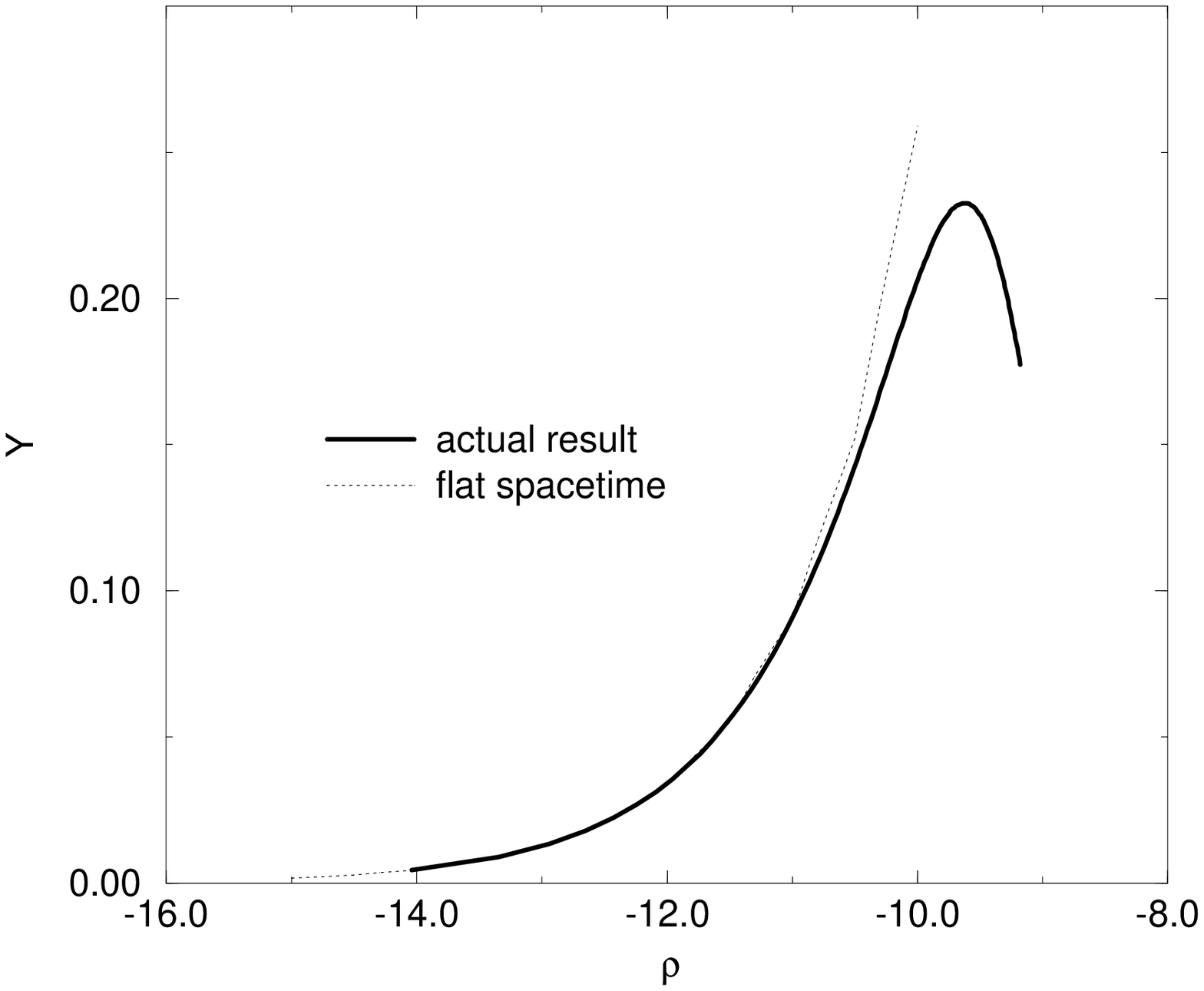}

${}^{}$\hspace{2cm}\epsfxsize=250pt \epsfbox{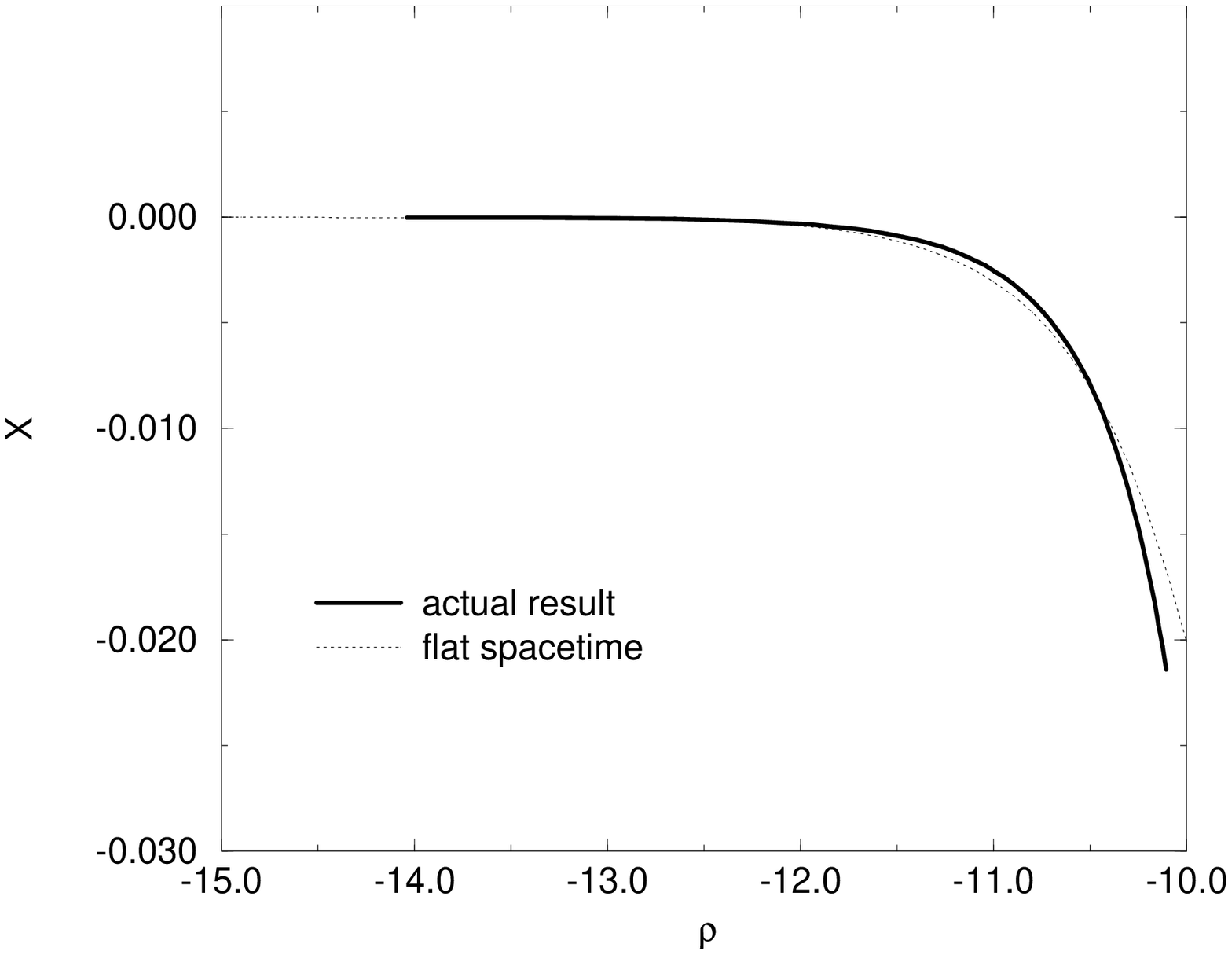}
\caption{
The solutions $X(\rho)$, $Y(\rho)$ found by Choptuik for
$\tau=-7.87488$ are compared to the quiescent region flat spacetime
solutions.  Parameters ${\cal A}=0.324, \bar{\delta}=0.372$, have been
used in the flat spacetime solution given in the text. }
\end{figure}

But at $\mu=0$, the solution is bounded as $\rho\rightarrow-\infty$,
so $F(\xi)$ and $G(\xi)$ must respectively have the form
$e^{-\xi/2}\tilde{F}(\xi), e^{\xi/2}\tilde{G}(\xi)$, where
$\tilde{F}(\xi),\tilde{G}(\xi)$ are bounded as
$\xi\rightarrow-\infty$. In the quiescent region, then, $x$ and $y$
have the form
\begin{eqnarray}
y&=&e^{\rho-\tau}\tilde{F}(\tau)\label{yform}\\ 
x&=&e^{\tau-\rho}\tilde{G}(\tau)\ .
\end{eqnarray}
At constant $\tau$, in the quiescent region, the scalar field
variables vanish as $\rho\rightarrow-\infty$.  This is satisfied by the above
form for $y$, but for $x$ not to diverge, $\tilde{G}$ must be zero.
This ``first approximation'' to $x$ and $y$ shows why
$X/Y\rightarrow0$ as $\rho\rightarrow-\infty$, and why the phase space
plots have vertical spikes corresponding to the
$\rho\rightarrow-\infty$ end of the plot. We can find a better
solution for $x$ by substituting the above solution for $y$ into the
left hand side of (\ref{eq2}), and by treating $a$ and $\alpha$ as constants.
This leads to
\begin{equation}\label{xform}
x=e^{-2\mu}\tilde{H}(\mu+\rho)\ ,
\end{equation}
where $\tilde{H}$ is given by
\begin{equation}
\tilde{H}(\xi)=-\frac{a}{3\alpha}\left[ \tilde{F}'(\xi)- 
\tilde{F}(\xi)\right]\ ,
\end{equation}
{}From the results of Choptuik, such as those shown in Figs.\ 1 and 2, it
is clear that the solutions in the quiescent region oscillate in
$\tau$, with a period of $2\pi/\omega$. The function $\tilde{F}$,
then, must have this periodic property.

In the above, we have considered $a$ and $\alpha$ to be constant, so
the solutions in (\ref{yform}) and (\ref{xform}), must be flat
spacetime solutions. To see this explicitly, we use the flat spacetime
forms of the solution in (\ref{flatY}) and (\ref{flatX}). Since $Y$ is
finite as $r\rightarrow0$ we must have $f(\eta)=-g(\eta)$. In the
limit of small $r$ this gives us
\begin{equation}
Y=-2rf''(t)+{\cal O}(r^2)\ .
\end{equation}
By comparison with (\ref{yform}) we have that
$f''(t)=-\tilde{F}(\tau)/2t$, where $\tilde{F}$ is a periodic function
with period $2\pi/\omega$. We illustrate the agreement of the Choptuik
results with such a flat spacetime solution, by choosing
\begin{equation}\label{fform}
f(\eta)={\cal A}\eta\ln{\eta}\sin{(\omega\ln{\eta}+\bar{\delta)}}.
\end{equation}
Figure 7 shows the agreement, in the quiescent region, between the
Choptuik results and the flat spacetime results of
(\ref{eq1}),(\ref{eq2}),(\ref{yform}), at $\tau=-7.87488$. In
(\ref{yform}) the parameter values ${\cal A}=0.324, \bar{\delta}=0.372$,
have been used.

\section{Discussion}

In the previous section we introduced two approximate solutions that
describe, with reasonable accuracy, the behavior of the Choptuik
spacetime and we showed that these solutions were equivalent to flat
spacetime solutions. Neither of these flat spacetime solutions, by
itself, provides an acceptable solution in the full range of the
spacetime manifold in which the Choptuik solution is found. From this
point of view, the essential role of general relativity, and the
nonlinearities it introduces into the equations, is to provide a match
of the two solutions across the transition edge. We give here a very
rough version of such a match. This is done to illustrate orders of
magnitudes, but it is amusing (and coincidental) that this approximate
match ends up giving a fairly accurate estimate of the period of the
Choptuik echoing.

We know from the flat spacetime approximation in the echoing region that,
\begin{equation}
1+{1\over \omega^2} = a^2\ .
\label{omegaa}
\end{equation}
We now adopt the viewpoint that $\omega$ is fixed by the way in which
$a$ increases across the transition edge. The requirement of
elementary flatness fixes $a$ to be unity at at $r=0$. Since $X$ and
$Y$ are small in the quiescent region, we see from (\ref{meq}), here rewritten
as 
\begin{equation}\label{rewrite}
({m\over r})_{,\rho} + {m\over r} =X^2+Y^2\ ,
\end{equation}
that $m/r$ must remain very small in the quiescent region, since $X$
and $Y$ are small. It follows that $a$ must be very close to unity in
the quiescent region, and this is verified by Choptuik's results in
Fig.\ 2.  Equation (\ref{rewrite}) also allows us to make a crude
estimate of how $a$ increases across the transition edge. From
(\ref{yform}) the discussion in the previous section we can
approximate $X\approx0$ and (with $\alpha$ taken to be approximately
unity) $Y\approx e^{-\mu}\tilde{F}(\tau)$. This approximation in
(\ref{rewrite}) gives us
\begin{equation}
({m\over r})_{,\rho} + {m\over r} =e^{2\rho}e^{-2\tau}\tilde{F}^2(\tau)\ .
\end{equation}
The initial condition on the solution is that $m/r=0$ at the origin
($\rho=-\infty$), so that the solution everywhere in the quiescent
region is approximately
\begin{equation}
{m\over r} = { 1 \over 3}\tilde{F}^2(\tau) e^{-2\mu}\ .
\end{equation}
We now assume that this solution gives a roughly correct answer at the
transition edge, $\mu=0$. Since $\tilde{F}$ is an oscillating
function, $m/r$ also oscillates (see Fig.\ 2), and a representative
value of $m/r$ is given by taking an average:
\begin{equation}\label{avg}
<{m\over r}> \approx\frac{1}{3}<\tilde{F}^2> \ .
\end{equation}
We now argue that $\tilde{F}$ must be of order unity. The transition
edge is defined as the location at which $X$ and $Y$ grow to order
unity. If this is to be at $\mu=0$ then, from $(\ref{yform})$ it is
clear that $\tilde{F}$ must be $\sim1$.  If, in fact, we take
$\tilde{F}$ to be a sinusoidal oscillation of amplitude unity, then
$<\tilde{F}^2>\approx1/2$, and we have from (\ref{avg}) that $<m/r>\approx1/6$.

As a consequence,
\begin{equation}
a^2 \sim 1 +<{2m \over r}> \sim {\textstyle 4 \over 3}\ ,
\end{equation}
which together with (\ref{omegaa}) implies,
\begin{equation}
\omega \sim \sqrt{3},\qquad \qquad \Delta \sim {2 \pi \over \sqrt{3}}
\sim 3.6
\end{equation}
Of course, we could equally well have concluded, for example, that
$<m/r>\approx0.1$ and $\Delta\sim2.8$. Our argument can only establish
that $\Delta$ is of order unity.  But this is enough to support our
suggestion that it is the matching provided by the nonlinearities that
picks out the form and the period of the Choptuik solution as the only
one that can bridge the two approximately flat-spacetime solutions.

\section*{Acknowledgments}

We wish to thank J\"urgen Ehlers, Bernd Schmidt and the Max Planck
Instit\"ut f\"ur Astrophysik at Garching for hospitality and support.
This work was supported in part by grants 
NSF-PHY-9423950, NSF-PHY-9396246, NSF-PHY-92-07225, NSF-PHY-95-07719,
research funds of the Pennsylvania State University, the University of
Utah, the Eberly Family research fund at PSU and PSU's Office for
Minority Faculty development. JP acknowledges support of the Alfred
P. Sloan foundation through a fellowship.

\end{document}